\documentclass[amstex,12pt,thmsa,sw20bams]{article}%
\usepackage{amsfonts}
\usepackage[utf8]{inputenc}
\usepackage{graphicx}
\usepackage[francais]{babel}
\usepackage{amsmath}
\usepackage{amssymb}%
\providecommand{\U}[1]{\protect\rule{.1in}{.1in}}
\ifx\pdfoutput\relax\let\pdfoutput=\undefined\fi
\newcount\msipdfoutput
\ifx\pdfoutput\undefined\else
\ifcase\pdfoutput\else
\ifx\paperwidth\undefined\else
\ifdim\paperheight=0pt\relax\else\pdfpageheight\paperheight\fi
\ifdim\paperwidth=0pt\relax\else\pdfpagewidth\paperwidth\fi
\fi\fi\fi
\begin{document}

\author{Choulakian V., Universit\'{e} de Moncton, Canada
\and vartan.choulakian@umoncton.ca}
\title{Notes on Correspondence Analysis of Power Transformed Data Sets}
\date{January 2023}
\maketitle

\begin{abstract}
We prospect for a clear simple picture on CA of power transformed or the
Box-Cox transformed data initiated since 2009 by Grenacre. We distinguish two
types of data sets: strictly positive and with zeros; we concentrate on the latter.



\end{abstract}

\section{\textbf{Introduction}}

Correspondence analysis (CA) and logratio analysis (LRA) are two popular
methods for the analysis and visualization of a two-way contingency table or a
compositional data set $\mathbf{N}=(n_{ij})$\textbf{ }of size $I\times J$ for
$i=1,...,I$ and $j=1,...,J.$ LRA is considered an ideal desirable method for
analysis-visualization of \textbf{N}, because it is row and column scales
invariant; but it can be applied only when $n_{ij}>0.$ While CA can be applied
to data when $n_{ij}\geq0$, but it is NOT row and column scale invariant
because it is dependent on its row and column marginals.

The central topic of this paper is Greenacre's (2010, Result 1), that we
rename it as theorem, and its application to $\mathbf{N}.$ Greenacre's Theorem
concerns the convergence of CA of $\mathbf{N}^{(\alpha)}=(n_{ij}^{\alpha})$
for $\alpha\in(0,1]$ as $\alpha\rightarrow0$ to uniformly weighted LRA of
$\mathbf{N}$ when $n_{ij}>0$. However, Greenacre (2011, 2022) attempts to show
its applicability - usefulness when $n_{ij}\geq0$; that is, when the data set
$\mathbf{N}$ contains zero valued cells. We aim to shed further light on this
latter situation.

This paper is organized as follows: Section 2 presents preliminaries; section
3 presents Greenacre's Theorem concerning CA of strictly positive data sets;
section 4 presents CA of nonnegative data sets having zero valued cells;
section 5 discusses three examples of real data sets with zero cells; finally
we conclude in section 6.

\subsection{Some references}

CA and LRA are based on three different principles: CA on Benz\'{e}cri's
\textit{distributional equivalence principle}, RC association models on Yule's
\textit{scale invariance principle}, and CoDA on Aitchison's
\textit{subcompositional coherence principle}. A recent discussion of these
three principles can be found in Choulakian et al. (2023).

Benz\'{e}cri (1973) is the reference book on CA. Beh and Lombardo (2014)
present a panoramic review of CA and its variants.

LRA includes two independently well developed methods: RC association models
for the analysis of contingency tables by Goodman (1979, 1981a, 1981b, 1991,
1996) and a set of compositional vectors (CoDA) by Aitchison (1986), see also
among others Greenacre (2018).

The relationship between CA and LRA has been discussed in many papers; see for
example, among others, Goodman (1996), Cuadras et al. (2006), Cuadras and
Cuadras (2015 ), Greenacre (2009, 2010), Beh and Lombardo (2022) and
Choulakian (2022).

\section{Preliminaries on analysis of contingency tables}

We consider a two-way contingency table $\mathbf{N}=(n_{ij})$ for $i=1,...,I$
and $j=1,...,J$, and $\mathbf{P=N/}n=(p_{ij})$ of size $I\times J$ the
associated correspondence matrix (probability table) of the contingency table
\textbf{N}. We define as usual $p_{i+}=\sum_{j=1}^{J}p_{ij}$ , $p_{+j}%
=\sum_{i=1}^{I}p_{ij},$ the vector $\mathbf{r=(}p_{i+})\in%
\mathbb{R}
^{I},$ the vector $\mathbf{c=(}p_{+j})\in%
\mathbb{R}
^{J}$, and $\mathbf{M}_{I}=Diag(\mathbf{r})$ the diagonal matrix having
diagonal elements $p_{i+},$ and similarly $\mathbf{M}_{J}=Diag(\mathbf{c}).$
We suppose that $\mathbf{M}_{I}$ and $\mathbf{M}_{J}$ are positive definite
metric matrices of size $I\times I$ and $J\times J$, respectively; this means
that the diagonal elements of $\mathbf{M}_{I}$ and $\mathbf{M}_{J}$ are
strictly positive.

\subsection{Independence of the row and column categories}

a) The $I$ row categories and the $J$ column categories are mutually
independent, then
\begin{equation}
\sigma_{ij}=p_{ij}-p_{i+}p_{+j}=0 \tag{1}%
\end{equation}
for $i=1,...,I$ and $j=1,...,J$, and, where $\sigma_{ij}$ is the residual
matrix of $p_{ij}$ with respect to the independence model $p_{i+}p_{+j}.$

\textbf{Remark 1}: The contingency table $\mathbf{N}=(n_{ij})$ can also be
represented (coded) as an indicator matrix $\mathbf{Z=}\left[  \mathbf{Z}%
_{I}\ \ \mathbf{Z}_{^{J}}\right]  =\left[  \mathbf{(}z_{\alpha i}%
)\ \ \mathbf{(}z_{\alpha j})\right]  \ $of size $n\times(I+J),$ where
$z_{\alpha i}=0$ if individual $\alpha$ does not have level $i$ of the row
variable, $z_{\alpha i}=1$ if individual $\alpha$ has level $i$ of the row
variable; $z_{\alpha j}=0$ if individual $\alpha$ does not have level $j$ of
the column variable, $z_{\alpha j}=1$ if individual $\alpha$ has level $j$ of
the column variable. Note that $\mathbf{N}=\mathbf{Z}_{^{I}}{}^{\prime
}\mathbf{Z}_{^{J}}\ $and $\sigma_{ij}=p_{ij}-p_{i+}p_{+j}$ is the covariance
between the $i$-th column of $\mathbf{Z}_{^{I}}$ and the $j$-th column of
$\mathbf{Z}_{^{J}}$.

b) The independence assumption $\sigma_{ij}=0$ can also be interpreted in
another way as%

\begin{align}
\Delta_{ij}  &  =(\frac{p_{ij}}{p_{i+}p_{+j}}-1)=0\tag{2}\\
&  =\frac{1}{p_{i+}}(\frac{p_{ij}}{p_{+j}}-p_{i+})\nonumber\\
&  =\frac{1}{p_{+j}}(\frac{p_{ij}}{p_{i+}}-p_{+j});\nonumber
\end{align}
this is the column and row homogeneity model. Benz\'{e}cri (1973, p.31) named
the conditional probability vector ($\frac{p_{ij}}{p_{+j}}$ for $i=1,...,I$
and $j$ fixed) the profile of the $j$-th column. He also referred to the
element of $\frac{p_{ij}}{p_{i+}p_{+j}}$ as\ the density function of the
probability measure $(p_{ij})$ with respect to the product measure
$p_{i+}p_{+j}$. The element $\frac{p_{ij}}{p_{i+}p_{+j}}$ is named Pearson
ratio in Goodman (1996) and Beh and Lombardo (2014, p.123).

c) A third way to represent the independence assumption $\sigma_{ij}=0$ and
the row and column homogeneity models $\Delta_{ij}=0$ is via the ($w_{i}^{R}$,
$w_{j}^{C})$ weighted loglinear formulation, equation (3), assuming $p_{ij}>0$
and defining $G_{ij}=\log(p_{ij}),$%

\begin{equation}
\lambda(p_{ij},w_{i}^{R},w_{j}^{C})=G_{ij}-G_{i+}-G_{+j}+G_{++}=0, \tag{3}%
\end{equation}
where $G_{i+}=\sum_{j=1}^{J}G_{ij}w_{j}^{C},$ $G_{+j}=\sum_{i=1}^{I}%
G_{ij}w_{i}^{R}$ and $G_{++}=\sum_{j=1}^{J}\sum_{i=1}^{I}G_{ij}w_{j}^{C}%
w_{i}^{R}$; $w_{j}^{C}>0$ and $w_{i}^{R}>0,$ satisfying $\sum_{j=1}^{J}%
w_{j}^{C}=\sum_{i=1}^{I}w_{i}^{R}\ =1,$ are a priori fixed or data dependent
probability weights. Two popular weights are marginal ($w_{i}^{R}%
=p_{i+},\ w_{j}^{C}=p_{+j})$ and uniform ($w_{i}^{R}=1/I,\ w_{j}^{C}=1/J).$
This is implicit in equation 7 in Goodman (1996) or equation 2.2.6 in Goodman
(1991); and explicit in Egozcue et al. (2015).

Equation (3) is equivalent to the logratios%
\[
\log(\frac{p_{ij}p_{i_{1}j_{1}}}{p_{ij_{1}}p_{i_{1}j}})=0\text{ \ \ for }i\neq
i_{1}\text{ and }j\neq j_{1},
\]
which Goodman (1979, equation 2.2) refers to as the \textquotedblright null
association\textquotedblright\ model.

Equation (3) is also equivalent to%
\[
p_{ij}=\frac{\exp(G_{i+})\exp(G_{+j})}{\exp(G_{++})},
\]
from which we deduce that : under the independence assumption the marginal row
probability vector ($p_{i+})$ is proportional to the vector of weighted
geometric means ($\exp(G_{i+}));$ and a similar property is true also for the
columns; see for instance Egozcue et al. (2015).

\subsection{Interaction factorization}

Suppose the independence-homogeneity-null association models are not true,
then each of the three equivalent model formulations (1), (2), (3) can be
generalized to explain the nonindependence-nonhomogeneity-association, named
interaction, among the $I$ rows and the $J$ columns by adding $k$ bilinear
terms, where $k=rank(\mathbf{N)-}1$.

We designate any one of the interaction indices (1), (2), (3) by $\tau_{ij}.$

Benz\'{e}cri (1973, Vol.2, p. 31-32) emphasized the importance of row and
column weights or metrics in multidimensional data analysis; this is the
reason in the French data analysis circles any study starts with a triplet
$\mathbf{(X,M}_{I},\mathbf{M}_{J}\mathbf{)}$, where \textbf{X} represents the
data set, $\mathbf{M}_{I}=(Diag(m_{i}^{r}))$ is the metric matrix defined on
the rows and $\mathbf{M}_{J}=(Diag(m_{j}^{c}))$ the metric matrix defined on
the columns. We follow the same procedure but $\mathbf{X}$ is the preprocessed
data, where:

a) In covariance analysis, $\mathbf{X}=(\tau_{ij}\mathbf{)=(}IJ\sigma_{ij})$ and

$\ \ \ \ \ \ \ \ \ \ (\mathbf{M}_{I},\mathbf{M}_{J})=(Diag(1/I),Diag(1/J)).$

b) In CA, $\mathbf{X}=(\tau_{ij}\mathbf{)=(}\Delta_{ij})$ and $(\mathbf{M}%
_{I},\mathbf{M}_{J})=(Diag(p_{i+}),Diag(p_{+j}));$

c) In LRA, $\mathbf{X}=(\tau_{ij}\mathbf{)=(}\lambda(p_{ij},w_{i}^{R}%
,w_{j}^{C}))$ and $(\mathbf{M}_{I},\mathbf{M}_{J})=(Diag(w_{i}^{R}%
),Diag(w_{j}^{C}))$ with $\sum_{j=1}^{J}w_{j}^{C}=\sum_{i=1}^{I}w_{i}^{R}=1.$

Note that in a) we have multiplied $\sigma_{ij}$ by $IJ$ to unify
theoretically the covariance analysis with the CA analysis, see the subsection
on double centering.

We factorize the interactions in (1), (2) and (3) by singular value
decomposition (SVD) or taxicab SVD (TSVD) as%
\begin{equation}
\tau_{ij}=\sum_{\alpha=1}^{k}f_{\alpha}(i)g_{\alpha}(j)/\delta_{\alpha},
\tag{4}%
\end{equation}

where $f_{\alpha}(i)$ is the $i$th row principal coordinate and $g_{\alpha
}(j)$ is the $j$th column principal coordinate along the $\alpha$th principal
dimension. Also $\delta_{\alpha}$ is the dispersion measure of the $\alpha$th
principal axis.

\textbf{Remark 2:}

a) In the SVD case the parameters $(f_{\alpha}(i),g_{\alpha}(j),\delta
_{\alpha})$ satisfy the conditions: for $\alpha,\beta=1,...,k$
\[
\delta_{\alpha}^{2}=\sum_{i=1}^{I}f_{\alpha}^{\ \ 2}(i)m_{i}^{r}=\sum
_{j=1}^{J}g_{\alpha}^{2}(j)m_{j}^{c}%
\]%
\[
0=\sum_{i=1}^{I}f_{\alpha}(i)m_{i}^{r}=\sum_{j=1}^{J}g_{\alpha}(j)m_{j}^{c}%
\]%
\[
0=\sum_{i=1}^{I}f_{\alpha}(i)f_{\beta}(i)m_{i}^{r}=\sum_{j=1}^{J}g_{\alpha
}(j)g_{\beta}(j)m_{j}^{c}\text{\ \ for }\alpha\neq\beta.
\]
b) In the TSVD case the parameters $(f_{\alpha}(i),g_{\alpha}(j),\delta
_{\alpha})$ satisfy the conditions: for $\alpha,\beta=1,...,k$
\[
\delta_{\alpha}=\sum_{i=1}^{I}|f_{\alpha}^{\ }|(i)m_{i}^{r}=\sum_{j=1}%
^{J}|g_{\alpha}(j)|m_{j}^{c}%
\]%
\[
0=\sum_{i=1}^{I}f_{\alpha}(i)m_{i}^{r}=\sum_{j=1}^{J}g_{\alpha}(j)m_{j}^{c}%
\]%
\[
0=\sum_{i=1}^{I}f_{\alpha}(i)\ sign(f_{\beta}(i))m_{i}^{r}=\sum_{j=1}%
^{J}g_{\alpha}(j)\ sign(g_{\beta}(j))m_{j}^{c}\text{\ \ for }\alpha>\beta.
\]
A description of TSVD can be found in Choulakian (2006, 2016).\bigskip

\textbf{Remark 3}

a) In the case $(\tau_{ij}\mathbf{)=(}IJ\sigma_{ij})$, the bilinear
decomposition (4) is also named interbattery analysis first proposed by Tucker
(1958); later on, Tenenhaus and Augendre (1996) reintroduced it within
correspondence analysis circles, where they showed that the Tucker
decomposition by SVD produced on some correspondence tables more interesting
(interpretable) structure than CA.

b) In the case $(\tau_{ij}\mathbf{)=(}\Delta_{ij})$, the CA decomposition has
many interpretations. Essentially, for data analysis, purposes Benz\'{e}cri
(1973) interpreted it as weighted principal components analysis of row and
column profiles. Another useful interpretation of CA, comparable to Tucker
interbattery analysis, is Hotelling(1936)'s canonical correlation analysis,
see Lancaster (1958) and Goodman (1991, 1996).

c) In the case $(\tau_{ij}\mathbf{)=(}\lambda(p_{ij},w_{i}^{R},w_{j}^{C}))$
and $(\mathbf{M}_{I},\mathbf{M}_{J})=(Diag(w_{i}^{R}),Diag(w_{j}^{C})),$ where
$(w_{i}^{R},w_{j}^{C})$ are prespecified; we note this case by TLRA or LRA;
for an example of general prespecified weights see Egozcue \ and
Pawlowsky-Glahn (2016). For the important particular case $\lambda
(p_{ij},w_{i}^{R},w_{j}^{C})=\lambda(p_{ij},1/I,1/J),$ we get
uniformly-weighted (or taxicab) logratio analysis uwLRA (or uwTLRA). \ 

d) We are concerned with the property of scale dependence or independence of
the three interaction indices (1), (2) and (3). The three indices depend on
$p_{ij}$ where $p_{ij}=n_{ij}/\sum_{i,j}n_{ij}.$ To emphasize this dependence,
we express any one of of the three interaction indices by $\tau_{ij}%
(n_{ij})=\tau(p_{ij},m_{i}^{R},m_{j}^{C}).$ Following Yule (1912), we state
the following\bigskip

\textbf{Definition 1}: An interaction index $\tau_{ij}(n_{ij})$ is scale
invariant if $\tau_{ij}(n_{ij})=\tau_{ij}(a_{i}n_{ij}b_{j})$ for arbitrary
scales $a_{i}>0$ and $b_{j}>0$.\bigskip

It is evident that neither of the three indices with data dependent marginal
weights $(p_{i+},p_{+j})$ are scale invariant.

Concerning the association index (3) we have the following lemma, see
Choulakian (2022) or Choulakian et al. (2023).\bigskip

\textbf{Lemma 1}: a) The association index (3) with prespecified weights
($w_{i}^{R},w_{j}^{C})$ is scale invariant. That is,
\[
\lambda(n_{ij},w_{i}^{R},w_{j}^{C})=\lambda(a_{i}n_{ij}b_{j}/n^{\ast}%
,w_{i}^{R},w_{j}^{C}),
\]
for arbitrary scales $a_{i}>0$ and $b_{j}>0$ and $n^{\ast}=\sum_{i,j}%
a_{i}n_{ij}b_{j}.$

b) To a first-order approximation, $\lambda(n_{ij},w_{i}^{R},w_{j}^{C}%
)\approx(\frac{p_{ij}}{w_{j}^{C}w_{i}^{R}}-\frac{p_{i+}}{w_{i}^{R}}%
-\frac{p_{+j}}{w_{j}^{C}}+1).$

\subsection{Double centering}

The interaction matrix $(\tau_{ij}\mathbf{)}$ is double centered, that is:%
\begin{align}
0  &  =\sum_{i=1}^{I}m_{i}^{r}m_{j}^{c}\tau_{ij}\nonumber\\
&  =\sum_{j=1}^{J}m_{i}^{r}m_{j}^{c}\tau_{ij}. \tag{5}%
\end{align}

According to Tukey (1977, chapter 10), there are two kinds of double centering
\textit{row-PLUS-column} and \textit{row-TIMES-column}; we name them additive
and multiplicative, that we describe.

We consider the triplet $\mathbf{(Y,M}_{I},\mathbf{M}_{J}\mathbf{)}$, where
$Y_{ij}=h(p_{ij})$ where $h$ is a function and $(\mathbf{M}_{I},\mathbf{M}%
_{J})=(Diag(m_{i}^{r}),Diag(m_{j}^{c})).$ We compute the three means, two
marginals and the total:$\ Y_{i+}=\sum_{j=1}^{J}Y_{ij}m_{j}^{c},$ $Y_{+j}%
=\sum_{i=1}^{I}Y_{ij}m_{i}^{r}$ and $Y_{++}=\sum_{j=1}^{J}\sum_{i=1}^{I}%
Y_{ij}m_{i}^{r}m_{j}^{c}.$

The multiplicative double centering is
\begin{equation}
\tau_{ij}=Y_{ij}-\frac{Y_{i+}Y_{+j}}{Y_{++}}, \tag{6}%
\end{equation}
where rank($\tau_{ij})=\ $rank$(Y_{ij})-1.$

The additive double centering is
\begin{equation}
\tau_{ij}=Y_{ij}-Y_{i+}-Y_{+j}+Y_{++}, \tag{7}%
\end{equation}
where rank($\tau_{ij})=\ $rank$(Y_{ij})-1$ or rank$(Y_{ij})-2.$

An evident difference between the two types of double centering is that (7) is
invariant to a row or column additive constants; Definition 1 and Lemma1a
become a corollary to this fact.

We consider two functional forms of $Y_{ij}=h(p_{ij}).$

a) $Y_{ij}=h(p_{ij})=\frac{p_{ij}}{m_{i}^{r}m_{j}^{c}}$ is the density
function of the joint probability measure $p_{ij}$ with respect to the product
measure $m_{i}^{r}m_{j}^{c}.$ First, the covariance interaction $(\tau
_{ij}\mathbf{)=(}IJ\sigma_{ij})$ is obtained from (6), when $m_{i}^{r}=1/I$
and $m_{j}^{c}=1/J.$ So the rank$\mathbf{(}IJ\sigma_{ij})=\ $rank($IJp_{ij}%
)-1.$ Second, the CA interaction $(\tau_{ij}\mathbf{)=(}\Delta_{ij})$ is
obtained either from (6) or from (7), when $m_{i}^{r}=p_{i+}$ and $m_{j}%
^{c}=p_{+j}.$ So the rank$\mathbf{(}\Delta_{ij})=\ $rank($\frac{p_{ij}}%
{p_{i+}p_{+j}})-1.$ Third, the right-hand side of Lemma 1b is obtained from
(7) when $m_{i}^{r}=w_{i}^{R}$ and $m_{j}^{c}=w_{j}^{C}.$

b) $Y_{ij}=h(p_{ij})=\log p_{ij}$ for $p_{ij}>0.$ First, the log interaction
$(\tau_{ij}\mathbf{)=(}\lambda(p_{ij},w_{i}^{R},w_{j}^{C}))$ is obtained from
(7), when $\ m_{i}^{r}=w_{i}^{R}$ and $m_{j}^{c}=w_{j}^{C}.$ So the rank
$\mathbf{(}\lambda(p_{ij},w_{i}^{R},w_{j}^{C}))=$ rank($\log p_{ij})-1,$ as in
Goodman (1991, Table 11); or rank($\log p_{ij})-2,$ as in Goodman (1991, Table
10). Second, to our knowledge the log interaction obtained from (6) has not
been applied yet, most probably due to the fact that it is not row and column
scale invariant; see also subsection 3.1.

Choulakian et al. (2023) used the quality of the sign of the residuals index,
QSR, to choose an optimal case among the different cases mentioned above.

\section{CA of power transformed strictly positive data}

Here, we state Greenacre's Theorem and provide a mathematical proof in the appendix.

Let ($p_{ij}^{(\alpha)}=\frac{n_{ij}^{\alpha}}{\sum_{i,j}n_{ij}^{\alpha}})$
for $i=1,...,I$ and \ $j=1,...,J\ $be the correspondence table of the power
transformed strictly positive data, and ($\Delta_{ij}^{(\alpha)}=\frac
{p_{ij}^{(\alpha)}-p_{i+}^{(\alpha)}p_{+j}^{(\alpha)}}{p_{i+}^{(\alpha)}%
p_{+j}^{(\alpha)}})$ the CA interaction matrix.

\textbf{Theorem }(Greenacre (2010, Result 1)

Under the assumption $n_{ij}>0,$ for $\alpha\rightarrow0$ and $\alpha>0$%
\begin{align*}
\Delta_{ij}^{(\alpha)}  &  =\frac{p_{ij}^{(\alpha)}-p_{i+}^{(\alpha)}%
p_{+j}^{(\alpha)}}{p_{i+}^{(\alpha)}p_{+j}^{(\alpha)}}\\
&  \approx\alpha\lambda(p_{ij},w_{i}^{R}=1/I,w_{j}^{C}=1/J)\\
&  =\alpha(\log p_{ij}+\frac{1}{IJ}\sum_{i,j}\log p_{ij}-\frac{1}{I}\sum
_{i}\log p_{ij}-\frac{1}{J}\sum_{j}\log p_{ij})
\end{align*}

Furthermore, $p_{i+}^{(\alpha)}\approx1/I$ for $i=1,...,I$ and $p_{+j}%
^{(\alpha)}\approx1/J$\ for \ $j=1,...,J.\bigskip$

A way to see the theorem is to look at the following sequence of
approximations using Lemma1b, and, the fact that $p_{i+}^{(\alpha)}\approx1/I$
for $i=1,...,I$ and $p_{+j}^{(\alpha)}\approx1/J$\ for \ $j=1,...,J$ as
$\alpha\rightarrow0.$%

\begin{align*}
\lambda(p_{ij}^{(\alpha)},w_{i}^{R}  &  =1/I,w_{j}^{C}=1/J)=\alpha
\lambda(p_{ij},w_{i}^{R}=1/I,w_{j}^{C}=1/J)\\
&  \approx(IJp_{ij}^{(\alpha)}+1-Ip_{i+}^{(\alpha)}-Jp_{+j}^{(\alpha)})\\
&  \approx(IJp_{ij}^{(\alpha)}-1)\\
&  \approx\Delta_{ij}^{(\alpha)}.
\end{align*}

\subsection{ Remarks}

a) The Theorem is interesting, but not useful in empirical contexts, because
one can directly compute the log interactions $\lambda(p_{ij},w_{i}%
^{R}=1/I,w_{j}^{C}=1/J).$

b) In the above Theorem, the assumption $n_{ij}>0$ is fundamental: if
$\alpha=0,$ then $n_{ij}^{\alpha}=n_{ij}^{0}=1$ for all $(i,j);$ that is, the
rank of the matrix ($n_{ij}^{\alpha}=n_{ij}^{0}=1)$ is 1; thus both sides of
the equation in the Theorem are zero.

c) In the proof of the above Theorem in the appendix, one sees that
\[
\lim_{\alpha\rightarrow0}\frac{\Delta_{ij}^{(\alpha)}}{\alpha}=\lambda
(p_{ij},w_{i}^{R}=1/I,w_{j}^{C}=1/J).
\]

d) Assume $n_{ij}\geq1;$ then another power transformation is the logarithmic
transformation, also known as the Box-Cox transformation, $\mathbf{L}%
=(logn_{ij}=\lim_{\alpha\rightarrow0}\frac{1}{\alpha}(n_{ij}^{\alpha}-1))$. We
have not seen any application of CA to \textbf{L}, see the subsection 2.3 on
double centering. Note that the assumption in Greenacre's Theorem $n_{ij}>0$
is weaker than the assumption $n_{ij}\geq1$ for the use of the log
transformation. Goodman's RC model, based on the log transformation, has been
developed for the analysis of contingency tables (tables with strictly
positive counts), and not for compositional data where often data in \% have
strictly positive values less than 1, besides having positive values larger
than 1; see for instance, the archeological compositional CUPS data in
Greenacre and Lewi (2008).

e) There is some kind of kinship between Greenacre's Theorem and Goodman's
marginal free CA (mfCA), see Goodman (1996, equation (46)). In mfCA of a
probability table $\mathbf{P=(}p_{ij})$ with row and column marginals
($p_{i+})$ and $(p_{+j})$ respectively, CA is applied to a matrix
$\mathbf{Q}=(q_{ij}=a_{i}p_{ij}b_{j})$ related to \textbf{P} in two steps:

Step 1: by the scale invariance property of the log interaction index, see
Lemma1a, under the assumption $n_{ij}>0$ there exists a unique probability
matrix $\mathbf{Q=}(q_{ij})$ which is related to $\mathbf{P=(}p_{ij})$ via the
strictly positive scales $(a_{i},b_{j}),$ that keeps Yule's association
between the $i$-th row and the $j$-th column unchanged; that is,
\[
\lambda(p_{ij},w_{i}^{R}=1/I,w_{j}^{C}=1/J)=\lambda(q_{ij}=a_{i}p_{ij}%
b_{j},w_{i}^{R}=q_{i+}=1/I,w_{j}^{C}=q_{+j}=1/J).
\]
Note that \textbf{Q} has uniform row and column marginal weights similar to
($p_{ij}^{(\alpha)})$.

Step 2, mfCA of \textbf{P} is CA representation of \textbf{Q}%
\[
IJq_{ij}-1=\sum_{\alpha=1}^{k}f_{\alpha}(i)g_{\alpha}(j)/\delta_{\alpha}.
\]

For further details, see Choulakian (2022).

\section{CA of power transformed data set having zero valued cells}

It is quite common that large compositional data sets or contingency tables
have zero valued cells. For this case, let $m$ be the number of zero valued
cells; so $IJ-m$ is the number of strictly positive valued cells in the table.
We see that for such tables the log transformation discussed in Remark b is
not valid, while the simple power transformation $n_{ij}^{\alpha}$ for
$\alpha\in\lbrack0,1]$ is valid. That is, the matrix $(lim\ n_{ij}^{\alpha})$
as $\alpha\rightarrow0$ converges to an indicator matrix $\mathbf{Z}%
=(z_{ij}),$ where $z_{ij}=1$ if $n_{ij}>0$ and $z_{ij}=0$ if $n_{ij}=0.$ A
comparison of this case with the observation in Remark a) under the assumption
$n_{ij}>0$ is insightful, where the indicator matrix becomes $\mathbf{Z}%
=(z_{ij}=1)$ of rank 1$.$ Here, it is insightful to consider two distinct
cases: Sparse tables containing multiple zeros in at least two nonproportional
rows or nonproportional columns, and tables with exactly one column (or row)
having at least one zero valued cell. In the latter case the zero entries
exhibit a dominating influence very similar to the cases of a heavyweight cell
and a heavyweight column (or row) discussed by Benz\'{e}cri (1979) and\ Lebart
(1979) in CA, and by Choulakian (2008) in TCA.

\subsection{Tables with one column having m zero-valued cells}

Here we discuss the case where there are $m$ for $m=1,...,I-1$ zero valued
cells in one column of the indicator matrix $\mathbf{Z}$ of size $I\times J.$
By Benz\'{e}cri's principle of distributional equivalence property which
states that in CA (or TCA) proportional rows or columns can be merged, CA (or
TCA) of \textbf{Z} is identical to CA (or TCA) of the $2\times2$ contingency
table
\begin{equation}
\mathbf{R}=\left[
\begin{tabular}
[c]{ll}%
$0$ & $m(J-1)$\\
$(I-m)$ & $(I-m)(J-1)$%
\end{tabular}
\right]  , \tag{8}%
\end{equation}
which has only one CA (or TCA) principal dimension.\bigskip

\textbf{Lemma 2}

a) In CA of \textbf{R} the dispersion, named inertia, $\rho_{1}^{2}=\frac
{m}{IJ.}.$

b) In TCA of \textbf{R} the taxicab dispersion $\delta_{1}=\frac
{4m(J-1)(I-m)}{(IJ-m)^{2}}.$

Proof of a): By equation 2.1.4 in Goodman (1996), for a $2\times2$ contingency
table the \textquotedblright ninindependence\textquotedblright\ based on the
correlation coefficient is
\begin{equation}
\rho_{1}^{2}=\frac{(p_{11}p_{22}-p_{12}p_{21})^{2}}{p_{1+}p_{2+}p_{+1}p_{+2}}.
\tag{9}%
\end{equation}
By replacing the probability values of the elements of \textbf{R }in (9), we
get%
\begin{align*}
\rho_{1}^{2}  &  =\frac{m^{2}(J-1)^{2}(I-m)^{2}}{(I-m)m(J-1)(J-1)I(I-m)J}\\
&  =\frac{m}{IJ}.
\end{align*}

Proof of b): We have to calculate the cross-covariance values $\sigma
_{ij}=p_{ij}-p_{i+}p_{+j}$ for $i=1,2$ and $j=1,2$ of \textbf{R}
\begin{align}
\mathbf{\Sigma}  &  =\left[
\begin{tabular}
[c]{ll}%
$\sigma_{11}$ & $\sigma_{12}$\\
$\sigma_{21}$ & $\sigma_{22}$%
\end{tabular}
\right] \nonumber\\
&  =\left[
\begin{tabular}
[c]{ll}%
$\sigma_{11}$ & $-\sigma_{11}$\\
$-\sigma_{11}$ & $\sigma_{11}$%
\end{tabular}
\right]  , \tag{10}%
\end{align}
because ($\sigma_{ij}=p_{ij}-p_{i+}p_{+j})$ is row and column centered.

There is one principal axis $\mathbf{u}_{1}^{\prime}=(1\ -1).$ So,
\begin{align*}
\delta_{1}  &  =||\mathbf{\Sigma u}_{1}||_{1}\\
&  =4|\sigma_{11}|\\
&  =4\frac{m(J-1)(I-m)}{(IJ-m)^{2}}.
\end{align*}

\subsection{Sparse tables}

Suppose $\mathbf{Z}=(z_{ij})$ is an incidence matrix, a presence-absence data
set, where $z_{ij}=0$ means level $j$ is absent in the $i$-th individual,
$z_{ij}=1$ means level $j$ is present in the $i$-th individual. CA (or TCA) is
a popular method for the analysis of such tables, see for an example
Choulakian and Abou-Samra (2020). Putting $p_{ij}=z_{ij}/\sum_{i,j}z_{ij}$ and
supposing that the marginals $p_{+j}>0$ and $p_{i+}>0$, CA (or TCA) data
reconstruction formula is%
\begin{equation}
p_{ij}=p_{+j}p_{i+}(1+\sum_{\alpha=1}^{k}f_{\alpha}(i)g_{\alpha}%
(j)/\delta_{\alpha}). \tag{11}%
\end{equation}
Now suppose we apply uniformly weighted LRA (or TLRA); observing $\log
_{2}(z_{ij}+1)=z_{ij}$, then the data reconstruction formula becomes
\begin{equation}
p_{ij}=p_{+j}/I+p_{i+}/J-1/(IJ)+\sum_{\alpha=1}^{k}f_{\alpha}(i)g_{\alpha
}(j)/\delta_{\alpha}; \tag{12}%
\end{equation}
a familiar one known as a FANOVA ( factor analysis and analysis of variance),
see Mandel (1971).

To choose the \textquotedblright best\textquotedblright\ between (11) and (12)
we use the quality of the signs of the residuals index, (QSR), within Taxicab
framework, see Choulakian(2021) and Choulakian et al. (2023).

\section{Examples}

Here we analyze three publicly available contingency tables and provide
summary results. For the computations we use the two R packages \textit{ca}
and \textit{TaxicabCA}.

\subsection{Example 1: Author data set}

Greenacre and Lewi (2009) discussed the \textit{author} contingency table that
has one zero valued count. This data set is of size $12\times26$ and is
included in the correspondence analysis \textit{ca }in R package by Greenacre
et al. (2022).

We consider the power transformed data set $\mathbf{N}^{(\alpha)}%
=(n_{ij}^{\alpha})$ for $\alpha=10\symbol{94}(-4);$ this value of $\alpha$ is
used by Greenacre (2022).$\ $By the \textit{ca} in R package the first two
dispersion-inertia values of $\mathbf{N}^{(\alpha)}$ are: \ \ \
\[
0.003204,\ \ \ \ 0;
\]
while by Lemma2a, via the merged indicator matrix \textbf{R} of size
$2\times2,$ with $m=1,\ I=12$ and $J=26$, we get the value of the first
principal inertia: $1/(12\ast26)=0.003205$.

By the \textit{TaxicabCA} in R package the first two dispersion values of
$\mathbf{N}^{(\alpha)}$ are: \ \ \
\[
0.011369,\ \ \ 9e-06;
\]
while by Lemma2b, via the \textbf{R} matrix of size $2\times2,$ with
$m=1,\ I=12$ and $J=26$, we get the value of the first principal taxicab
dispersion: $0.011373$.

This example shows the dominant influence of one zero-valued cell in
$\mathbf{N}^{(\alpha)}=(n_{ij}^{\alpha})$ for $\alpha=10\symbol{94}(-4).$ The
next example shows that similar result is also obtained for multiple
zero-valued cells in one column.

\subsection{Example 2: RBGlass1 data set}

\textit{RBGlass1} is a compositional data set of size $105\times11$ that has
$m=26$ zeros in the variable Sb column. It is found in the R package
\textit{archdata} by Carlson and Roth (2022).

We consider the power transformed data set $\mathbf{N}^{(\alpha)}%
=(n_{ij}^{\alpha})$ for $\alpha=10\symbol{94}(-4).\ $By the \textit{ca} in R
package the first two dispersion-inertia values of $\mathbf{N}^{(\alpha)}$
are: \ \ \
\[
0.0225,\ 0;
\]
while by Lemma2a, via the merged indicator matrix \textbf{R} of size
$2\times2,$ with $m=26,\ I=105$ and $J=11$, we get the value of the first
principal inertia: $26/(11\ast105)=0.022511$.

By the TaxicabCA in R package the first two dispersion values of
$\mathbf{N}^{(\alpha)}$ are: \ \ \
\[
0.0645,\ \ \ 7e-06;
\]
while by Lemma2b, via the \textbf{R} matrix of size $2\times2,$ with
$m=26,\ I=105$ and $J=11$, we get the value of the first principal taxicab
dispersion: $0.0645$.

\subsection{Example 3: Rodent abundance data}

We consider the \textit{rodent} data set of size 28 by 9 found in the R
package \textit{TaxicabCA}. This is an abundance data set of 9 species of rats
in 28 cities in California. Choulakian (2017) analyzed it by comparing the CA
and TCA maps; Choulakian (2021) showed that it has quasi-2-blocks diagonal
structure; furthermore Choulakian (2022) analyzed it by Goodman's
marginal-free CA and marginal-free TCA methods.

Let \textbf{N} be the original data set of size $28\times9$; the
\textit{apparent} percentage of zero counts in \textbf{N} is 66.27\%\textbf{.}
The function \textquotedblright\textit{CombineCollinearRowsCols}%
\textquotedblright\ in the package TaxicabCA in R merges the rows and the
columns of \textbf{N}, which are proportional; we see that the size of the
\textbf{Nmerged} is $21\times9$.\textbf{ }So within the CA framework the
\textit{real} percentage of zero counts in \textbf{N} is the percentage of
zero counts in \textbf{Nmerged}, which is\textbf{ }58.73\%. Similarly the
\textit{real} size of the indicator matrix \textbf{Z} is the size of the
\textbf{Zmerged}, which is $14\times9,$ whose \% of zeros is 60.32\%.

We consider the power transformed data set $\mathbf{N}^{(\alpha)}%
=(n_{ij}^{\alpha})$ for $\alpha=10\symbol{94}(-4)$. The \textit{ca} in R
package produced exactly the same maps applied to $\mathbf{N}^{(\alpha)}$ and
\textbf{Zmerged}. Similarly,\textbf{ }the \textit{TaxicabCA} in R package
produced exactly the same maps applied to $\mathbf{N}^{(\alpha)}$ and
\textbf{Zmerged.}

The four maps can be seen by applying the R code in the appendix.

\section{Conclusion}

We attempted to have a simple clear picture on CA of power transformed or the
Box-Cox transformed data initiated since 2009 by Grenacre. We distinguished
two types of data sets: strictly positive and with zeros. In particular, we
showed the dominant influence of the zero entries in the CA of power
transformed data when the power goes to zero; in this case the power
transformed data set becomes almost a 0-1 indicator matrix.

An alternative approach is to add a positive constant to any zero-valued cell,
then use the log transformation as in the development of (11); see, among
others, Lubbe et al. (2021) and Choulakian et al. (2023).\bigskip
\begin{verbatim}
Acknowledgements.
\end{verbatim}

Choulakian's research has been supported by NSERC of Canada.\bigskip
\begin{verbatim}
Appendix 1: The R code
\end{verbatim}

The execution of the R code will produce the numerical results and the maps
discussed in the paper. The R code uses the following three packages.

a) The \textit{ca} package, by Greenacre et al. (2022), does CA and produces
the CA map.

b) The \textit{TaxicabCA} package, by Allard and Choulakian (2019), does TCA
and produces the TCA map.

c) The \textit{archdata} package, by Carlson and Roth (2022) for the
\textit{RBGlasse1} dataset.

\# install packages

install.packages(c("ca", "TaxicabCA", "archdata"))

library(TaxicabCA)

library(ca)

library(archdata)

\#Choose a data set

dataMatrix = as.matrix(rodent)

dataMatrix
$<$%
- t(author)

data(RBGlass1)

dataMatrix
$<$%
- RBGlass1[, -1]

dim(dataMatrix)

\#Compute dataMerged and IndicatorM

dataMerged
$<$%
- CombineCollinearRowsCols(dataMatrix, rows = T, cols = T)

dim(dataMerged)

IndicatorM = 1-(dataMatrix == 0)

IndicMerged
$<$%
- CombineCollinearRowsCols(IndicatorM, rows = T, cols = T)

dim(IndicMerged)

\#Compute dataPowered and its marginals

alpha
$<$%
- 10\symbol{94}(-4)

dataPowered
$<$%
- dataMatrix\symbol{94}alpha

sum(dataPowered)

apply(dataPowered, 2, function(x) sum(x))

apply(dataPowered, 1, function(x) sum(x))

\#CA map of rodent dataset

plot(ca(dataMatrix))

plot(ca(dataPowered))

plot(ca(IndicatorM))

\# TCA maps

tca.Data
$<$%
- tca(dataMatrix, nAxes=2,algorithm = "exhaustive")

plot(tca.Data, axes = c(1, 2),labels.rc = c(2, 2))

tca.Data
$<$%
- tca(dataPowered, nAxes=2,algorithm = "exhaustive")

plot(tca.Data, axes = c(1, 2),labels.rc = c(2, 2))

tca.Data
$<$%
- tca(IndicatorM, nAxes=2,algorithm = "exhaustive")

plot(tca.Data, axes = c(1, 2),labels.rc = c(2, 2))\bigskip
\begin{verbatim}
Appendix 2:
\end{verbatim}

\textbf{Theorem} (Greenacre (2010, Result 1)

Under the assumption $n_{ij}>0,$
\begin{align*}
\Delta_{ij}^{(\alpha)}  &  =\frac{p_{ij}^{(\alpha)}-p_{i+}^{(\alpha)}%
p_{+j}^{(\alpha)}}{p_{i+}^{(\alpha)}p_{+j}^{(\alpha)}}\\
&  \approx\alpha\lambda(p_{ij},w_{i}^{R}=1/I,w_{j}^{C}=1/J)\\
&  =\alpha(\log p_{ij}+\frac{1}{IJ}\sum_{i,j}\log p_{ij}-\frac{1}{I}\sum
_{i}\log p_{ij}-\frac{1}{J}\sum_{j}\log p_{ij})
\end{align*}
for $\alpha\rightarrow0$ and $\alpha>0.$

Proof: By Maclaurin-Taylor series expansion, in the neighborhood of
$\alpha=0,$ $n_{ij}^{\alpha}=\exp(\alpha\log n_{ij})=1+\alpha\log
x+O(\alpha^{2}),$ where $r(\alpha)=O(h(\alpha))$ means $\lim_{\alpha
\rightarrow0}|\frac{r(\alpha)}{h(\alpha)}|=constant>0.$ So$\ $\ for
$i=1,...,I\ \ $and\ \ $j=1,...,J$%
\begin{align}
p_{ij}^{(\alpha)}  &  =\frac{n_{ij}^{\alpha}}{\sum_{i,j}n_{ij}^{\alpha}%
}\nonumber\\
&  =\frac{1+\alpha\log n_{ij}+O(\alpha^{2})}{IJ+\sum_{i,j}\alpha\log
n_{ij}+O(\alpha^{2})}\tag{A1}\\
&  =\frac{\left[  1+\alpha\log n_{ij}+O(\alpha^{2})\right]  \left[
IJ+\alpha\sum_{i,j}\log n_{ij}+O(\alpha^{2})\right]  }{\left[  IJ+\alpha
\sum_{i,j}\log n_{ij}+O(\alpha^{2})\right]  ^{2}}\nonumber\\
&  =\frac{IJ+\alpha IJ\log n_{ij}+\alpha\sum_{i,j}\log n_{ij}+O(\alpha^{2}%
)}{\left[  IJ+O(\alpha)\right]  ^{2}}. \tag{A2}%
\end{align}

By (5a) we have%

\begin{align}
p_{i+}^{(\alpha)}  &  =\sum_{j}p_{ij}^{(\alpha)}\nonumber\\
&  =\frac{J+\alpha\sum_{j}\log n_{ij}+O(\alpha^{2})}{IJ+O(\alpha)}; \tag{A3}%
\end{align}%
\begin{equation}
p_{+j}^{(\alpha)}=\frac{I+\alpha\sum_{i}\log n_{ij}+O(\alpha^{2})}%
{IJ+O(\alpha)}; \tag{A4}%
\end{equation}%
\[
p_{i+}^{(\alpha)}p_{+j}^{(\alpha)}=\frac{IJ+\alpha J\sum_{i}\log n_{ij}+\alpha
I\sum_{j}\log n_{ij}+O(\alpha^{2})}{\left[  IJ+O(\alpha)\right]  ^{2}}.
\]
By (A1) and (A4), we have%
\begin{align*}
\Delta_{ij}^{(\alpha)}  &  =\frac{p_{ij}^{(\alpha)}-p_{i+}^{(\alpha)}%
p_{+j}^{(\alpha)}}{p_{i+}^{(\alpha)}p_{+j}^{(\alpha)}}\\
&  =\frac{\left[  IJ+\alpha IJ\log n_{ij}+\alpha\sum_{i,j}\log n_{ij}%
+O(\alpha^{2})\right]  -\left[  IJ+\alpha J\sum_{i}\log n_{ij}+\alpha
I\sum_{j}\log n_{ij}+O(\alpha^{2})\right]  }{IJ+\alpha J\sum_{i}\log
n_{ij}+\alpha I\sum_{j}\log n_{ij}+O(\alpha^{2})}\\
&  =(\frac{\alpha IJ}{IJ})\frac{\log n_{ij}+\frac{1}{IJ}\sum_{i,j}\log
n_{ij}-\frac{1}{I}\sum_{i}\log n_{ij}-\frac{1}{J}\sum_{j}\log n_{ij}%
+O(\alpha)}{1+\alpha\frac{1}{I}\sum_{i}\log n_{ij}+\alpha\frac{1}{J}\sum
_{j}\log n_{ij}+O(\alpha^{2})}\\
&  =\alpha\frac{\lambda(p_{ij},w_{i}^{R}=1/I,w_{j}^{C}=1/J)+O(\alpha
)}{1+O(\alpha)};
\end{align*}

which is the required result by Lemma 1.

\textbf{Corollary}

By (A3) and (A4), we see that $p_{i+}^{(\alpha)}=1/I+O(\alpha)$ for
$i=1,...,I$ and $p_{+j}^{(\alpha)}=1/J+O(\alpha)$\ for \ $j=1,...,J.\bigskip$

\textbf{References}

Aitchison J (1986) \textit{The Statistical Analysis of Compositional Data}.
London: Chapman and Hall

Allard J, Choulakian V (2019) \textit{Package TaxicabCA in R}

Beh E, Lombardo R (2014) \textit{Correspondence Analysis: Theory, Practice and
New Strategies}. N.Y: Wiley

Beh E, Lombardo R (2022) Correspondence Analysis and the Cressie-Read
Divergence Statistic. National Institute for Applied Statistics Research ,
University of Wollongong, Australia, Working Paper 06-22

Benz\'{e}cri JP (1973)\ \textit{L'Analyse des Donn\'{e}es: Vol. 2: L'Analyse
des Correspondances}. Paris: Dunod

Carlson D.L, Roth G (2022) \textit{Package archdata in R}

Choulakian V (2006) Taxicab correspondence analysis. \textit{Psychometrika,}
71, 333-345

Choulakian V (2008) Taxicab correspondence analysis of contingency tables with
one heavyweight column. \textit{Psychometrika}, 73(2), 309-319

Choulakian V (2016) Matrix factorizations based on induced norms.
\textit{Statistics, Optimization and Information Computing}, 4, 1-14

Choulakian V (2017) Taxicab correspondence analysis of sparse contingency
tables. \textit{Italian Journal of Applied Statistics,} 29 (2-3), 153-179

Choulakian V (2021) Quantification of intrinsic quality of a principal
dimension in correspondence analysis and taxicab correspondence analysis.
Available on \textit{arXiv:2108.10685}

Choulakian V (2022) Some notes on Goodman's marginal-free correspondence
analysis. \textit{https://arxiv.org/pdf/2202.01620.pdf}

Choulakian V, Allard J, Mahdi S (2023) \ Taxicab correspondence analysis and
Taxicab logratio analysis: A comparison on contingency tables and
compositional data. To appear in \textit{Austrian Journal of Statistics}.

Cuadras CM, Cuadras D, Greenacre M (2006) A comparison of different methods
for representing categorical data. \textit{Communications in Statistics-Simul.
and Comp,} 35(2), 447-459

Cuadras CM, Cuadras D (2015 ) A unified approach for the multivariate analysis
of contingency tables. \textit{Open Journal of Statistics}, 5, 223-232

Egozcue JJ, Pawlowsky-Glahn V (2016) Changing the reference measure in the
simplex and its weighting effects. \textit{Austrian Journal of Statistics},
45(4), 25-44

Egozcue JJ, Pawlowsky-Glahn V, Templ M, Hron K (2015) Independence in
contingency tables using simplicial geometry. \textit{Communications in
Statistics - Theory and Methods}, 44:18, 3978-3996

Goodman LA (1979) Simple models for the analysis of association in
cross-classifications having ordered categories. \textit{Journal of the
American Statistical Association}, 74,537-55

Goodman LA (1981a) Association models and the bivariate normal for contingency
tables with ordered categories. \textit{Biometrika}, 68, 347-355

Goodman LA (1981b) Association models and canonical correlation in the
analysis of cross-classifications having ordered categories. \textit{Journal
of the American Statistical Association}, 76, 320-334

Goodman, LA (1991) Measures, models, and graphical displays in the analysis of
cross-classified data. \textit{Journal of the American Statistical
Association}, 86 (4), 1085-1111

Goodman LA (1996) A single general method for the analysis of cross-classified
data: Reconciliation and synthesis of some methods of Pearson, Yule, and
Fisher, and also some methods of correspondence analysis and association
analysis.\textit{\ Journal of the American Statistical Association}, 91, 408-428

Greenacre M (2009) Power transformations in correspondence analysis.
\textit{Computational Statistics \& Data Analysis,} 53(8), 3107-3116

Greenacre M (2010) Log-ratio analysis is a limiting case of correspondence
analysis. \textit{Mathematical Geosciences,} 42, 129-134

Greenacre M (2011) Measuring subcompositional incoherence.
\textit{Mathematical Geosciences,} 43, 681--93

Greenacre M (2022) The chi-square standardization, combined with Box-Cox
transformation, is a valid alternative to transforming to logratios in
compositional data analysis. Available at
\textit{https://arxiv.org/abs/2211.06755}

Greenacre M, Lewi P (2009) Distributional equivalence and subcompositional
coherence in the analysis of compositional data, contingency tables and
ratio-scale measurements. \textit{Journal of Classification}, 26, 29-54

Greenacre M, Nenadic O, Friendly M (2022) \textit{Package ca in R}

Hotelling H (1936) Relations between two sets of variables.
\textit{Biometrika} 28, 321-377

Lancaster HO (1958). The structure of bivariate distributions. \textit{Ann.
Math. Statist.} 29, 719--736

Mandel J (1971) A new analysis of variance model for non-additive data.
\textit{Technometrics}, 13(1), 1-18

Lubbe S, Filzmoser P, Templ M (2021) Comparison of zero replacement strategies
for compositional data with large numbers of zeros. \textit{Chem Intell Lab
Syst}, 210, 104248.

Tenenhaus M, Augendre H (1996) Analyse factorielle inter-batteries de Tucker
et analyse canonique aux moindres carr%
\'{}%
es partiels. In \textit{Recueil des r\'{e}sum\'{e}s des communications des
28`eme Journ\'{e}es de statistique}, 693-697

Tucker LR (1958) An inter-battery method of factor analysis.
\textit{Psychometrika}, 23, 111-136

Tukey JW (1977) \textit{Exploratory Data Analysis}. Addison-Wesley: Reading, Massachusetts

Yule GU (1912) On the methods of measuring association between two attributes.
\textit{JRSS}, 75, 579-642
\begin{verbatim}

\end{verbatim}

\end{document}